\documentclass[referee]{aa}
\usepackage{graphicx}
\usepackage{psfig}
\usepackage{amsfonts}

\begin{document}

\title{ $RRAB$ VARIABLES and METALLICITY }

\author{Z.Kadla\inst{1} }

\offprints{Z.Kadla, The Main Astronomical Observatory of the
Russian Academy of Sciences at Pulkovo, Saint-Petersburg, Russia.
email: kadla@gao.spb.ru}

 \institute{Main Astronomical Observatory of
the Russian Academy of Sciences at Pulkovo, 196140
Saint-Petersburg, Russia }

\date{ }

 \abstract

\maketitle

\begin{abstract} - A preliminary analysis of the $UBV$ light curves of $RR AB$
variables revealed that they contain information on the
metallicities of these stars. The computed parameters which
correlate with metallicity are given in the tables for 99 $RR AB$
stars including 12 variables without previously determined
metallicities.

\end{abstract}

\section{Introduction}
 Suntzeff, Kraft and Kinman (1994, SKK) determined
the metallicity of 84 bright $RR AB$ stars from spectra taken at
the Lick Observatory. The $[Fe/H]_{SKK}$ values are on the Zinn
and West (1994) scale. Blanco (1992) reduced Preston's $\Delta$S
values determined by various authors to a uniform system. The
resulting $\Delta$S for each variable is given in table 1 of the
paper. The reliable values are marked by an asterisk. Layden
(1994,L) observed the spectra of 302 $RR AB$ stars. The derived
abundances $[Fe/H]_{L}$ were calibrated to the Zinn and West
metallicity scale (1994). Jurcsik and Kovacs (1996) used the data
of the three above investigations for a Fourier analysis of $V$
light curves of $RR AB$ $Lyrae$ stars. They found a
period-phase-metallicity linear relation which is given in their
paper.

\section{ Photometric Data}
Light  curves  and  $UBV$  data  in  dependence on phase (phase
interval ph = 0.02) for $RR$ $Lyrae$ stars are given in the
catalogue "The Mean Light and Colour $(B-V)$ and $(U-B)$ Curves of
210 Field $RR$ $Lyrae$ Type Stars" compiled by  N. Nikolov, N.
Buchantsova and M. Frolov (1984, NBF). The authors reduced
available photometric observations of $RR$ $Lyrae$ variables to a
uniform system, the $UBV$ observations of Fitch et al (1966) being
used as standards.

\section{Analysis}

In the list compiled by Blanco (1992) data are given for 99 NBF
variables 62 of which have reliable determinations. For ten of
these stars $|[Fe/H]_{L} - [Fe/H]_{LS}|>$0.3. The following
dependence calculated for the remaining 52 was used for
determining $[Fe/H]_{LS}$ (r - correlation coefficient):

$[Fe/H]_{LS}$ = -0.291 - 0.178$\Delta$S; r=0.97.

 In SKK metallicities are given for 69 NBF variables.
The NBF $UBV$ photometric data were analysed in a search for a
possible correlation with metallicity $[Fe/H]_{L}$ using the
equations:

$V=a_{vu}+b_{VU}U; (VU);$    $U=a_{uv}+b_{UV}V; (UV);$

$B=a_{bu}+b_{BU}U; (BU);$    $U=a_{ub}+b_{UB}B; (UB).$

 The $V$ and $U$ and also $B$ and $U$ relations for the $RR AB$ star RR Cet
are shown in fig. 1.

 \begin{figure}[ht]
\centering{
\vbox{\psfig{figure=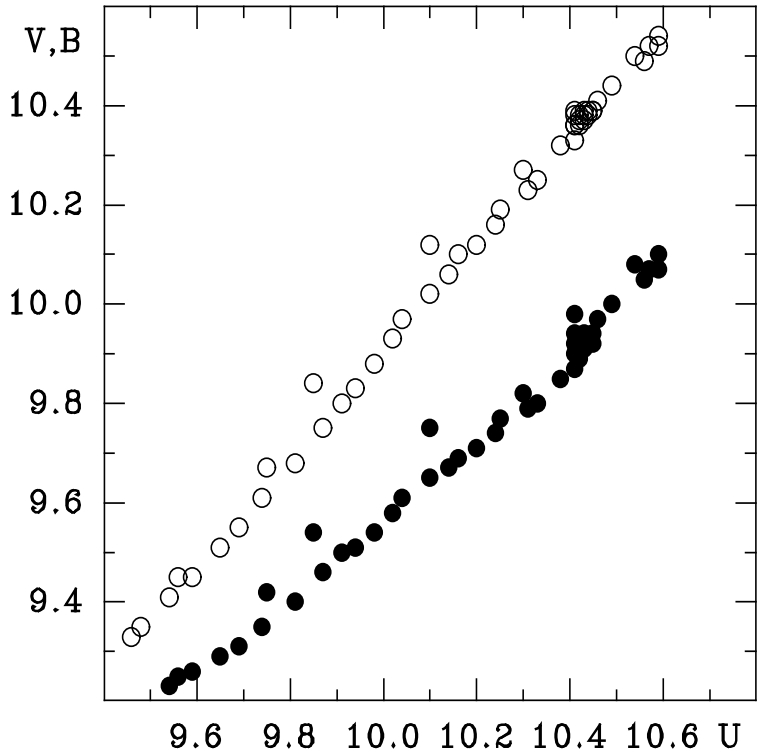,width=10.0cm,height=10.0cm}}\par}
\caption []{ RR Cet. The $V$ and $U$ $(\circ)$ and the $B$ and $U$
$(\bullet)$ relations. }
\end{figure}

The differences $([Fe/H]_{L}-[Fe/H]_{LVU})$,
$([Fe/H]_{L}-[Fe/H]_{LUV})$, $([Fe/H]_{L}-[Fe/H]_{LBU})$ and
$([Fe/H]_{L}-[Fe/H]_{LUB})$ were obtained for 29 NFB variables
with the number of measured phase intervals $nph>45$.

 The following relations were used in the final
solution ( n=29):

$[Fe/H]_{LVU} = 8.652 - 12.396b_{VU}$; r=-0.97;

 $[Fe/H]_{LUV} = -11.118 + 7.946b_{UV}$; r= 0.97;

 $[Fe/H]_{LBU} = 13.335 - 13.686b_{BU}$; r=-0.96;

 $[Fe/H]_{LUB} = -15.334 + 15.071b_{UB}$; r= 0.96.

In the three accompanying tables  the calculated metallicities
$[Fe/H]_{UBV}$ = $[Fe/H]_{LUBV}$  = $([Fe/H]_{LUV} + [Fe/H]_{LVU}
+ [Fe/H]_{LBU} + [Fe/H]_{LUB})/4$ are given for 95 $RR AB$ stars.
Altogether 21 including 11 without previous determinations, were
not measured by Layden (1994) (table 2).

The comparison of $[Fe/H]_{L}$ and $[Fe/H]_{UBV}$ values for 77
stars of table 1

$[Fe/H]_{UBV}$ = 0.010 + 0.971$[Fe/H]_{L}$ ; r=0.96;

is illustrated by fig.2.

 \begin{figure} [ht]
\centering{
\vbox{\psfig{figure=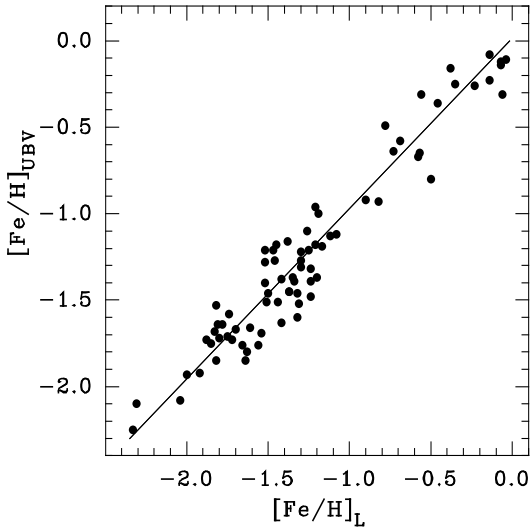,width=7.0cm,height=7.0cm}}\par}
\caption[]{ The comparison of the metallicity values $[Fe/H]_{L}$
and $[Fe/H]_{UBV}$. }
\end{figure}

Among the stars with sufficient data there are 8 with $nph > 45$
(table 3) the metallicities of which differ from those determined
by spectral analysis. In the tables data for period and $V$, $B$
and $U$ amplitudes are taken from NBF.

\begin{table}[h]
\caption[ ]{}
\begin{tabular}{lccccccccc}
\hline \noalign{\smallskip}
 Var & nph & Per & $A_{V}$& $A_{B}$& $ A_{U}$& $[Fe/H]_{L
}$&$[Fe/H]_{UBV}$&$[Fe/H]_{SKK}$&$[Fe/H]_{LS}$\\ \hline
\noalign{\smallskip}
 SW And &    26 &  0.442 & 0.93 & 1.26
& 1.38 & -0.38 & -0.16 & -0.41 & -0.22 \\ WY Ant &    38 &  0.574
& 0.91 & 1.21 & 1.38 & -1.66 & -1.76 &       &       \\ SW Aqr &
22 &  0.459 & 1.24 & 1.57 & 1.74 & -1.24 & -1.32 & -1.42 & -1.38
\\ TZ Aqr &    50 &  0.571 & 0.80 & 1.07 & 1.20 & -1.24 & -1.48 &
&       \\ BO Aqr &    49 &  0.694 & 1.04 & 1.36 & 1.52 & -1.80 &
-1.72 &       &       \\ CP Aqr &    50 &  0.463 & 1.26 & 1.64 &
1.75 & -0.90 & -0.92 & -0.77 &       \\ DN Aqr &    50 &  0.634 &
0.78 & 1.02 & 1.14 & -1.63 & -1.80 &       &       \\ AA Aql & 50
&  0.362 & 1.28 & 1.70 & 1.84 & -0.58 & -0.67 & -0.63 & -0.31
\\ V341 Aql&   50 &  0.578 & 1.21 & 1.56 & 1.68 & -1.37 & -1.45 &
-1.15 & -1.11 \\ SW Boo &    19 &  0.514 & 1.08 & 1.40 & 1.58 &
-1.12 & -1.13 &       &       \\ RW Cnc &    23 &  0.547 & 0.63 &
0.88 & 0.99 & -1.52 & -1.21 &       &       \\ SS Cnc &    17 &
0.367 & 1.18 & 1.63 & 1.75 & -0.07 &  0.12 & -0.56 & -0.61 \\ W
CVn &     38 &  0.552 & 0.81 & 1.11 & 1.24 & -1.21 & -0.96 & -1.17
& -1.36 \\ RR CVn &    19 &  0.559 & 0.93 & 1.27 & 1.47 & -1.08 &
-1.12 &       &       \\ RX CVn &    13 &  0.540 & 0.75 & 0.99 &
1.10 & -1.31 & -1.52 &       &       \\ RZ CVn &    25 &  0.567 &
0.89 & 1.20 & 1.40 & -1.92 & -1.92 & -1.81 & -1.04 \\ IU Car & 49
& 0.737 & 0.95 & 1.23 & 1.39 & -1.85 & -1.75 &       &       \\
V499 Cen &    50 &  0.521 & 1.17 & 1.53 & 1.68 & -1.56 & -1.76 & &
\\ RR Cet &    50 &  0.553 & 0.89 & 1.18 & 1.32 & -1.52 & -1.40 &
-1.55 & -1.41 \\ RX Cet &    29 &  0.574 & 0.76 & 1.01 & 1.14 &
-1.46 & -1.27 &       &       \\ V413 Cra &  50 &  0.589 & 0.67 &
0.90 & 1.00 & -1.21 & -1.18 &       &       \\ W Crt &     50 &
0.412 & 1.26 & 1.66 & 1.78 & -0.50 & -0.80 & -1.10 & -1.00 \\ X
Crt &     50 &  0.733 & 0.61 & 0.81 & 0.92 & -1.75 & -1.71 & &
\\ XZ  Cyg &   28 &  0.467 & 0.96 & 1.28 & 1.43 & -1.52 & -1.28 &
-1.39 & -1.39 \\ DM Cyg &    19 &  0.426 & 0.95 & 1.37 & 1.48 &
-0.14 & -0.23 & -0.57 & -0.59 \\ DX Del  &   50 &  0.473 & 0.70 &
0.97 & 1.04 & -0.56 & -0.31 & -0.50 & -0.49 \\ SU Dra & 20 & 0.660
& 0.98 & 1.29 & 1.43 & -1.74 & -1.58 & -1.79 & -1.96
\\ SW Dra &    16 &  0.570 & 0.92 & 1.21 & 1.37 & -1.24 & -1.39 &
-1.21 & -0.91 \\ RX Eri &    35 &  0.587 & 0.84 & 1.15 & 1.29 &
-1.30 & -1.27 & -1.43 & -1.98 \\ SV Eri &    43 &  0.714 & 0.59 &
0.80 & 0.92 & -2.04 & -2.08 & -1.70 & -1.84 \\ BB Eri &    39 &
0.570 & 0.84 & 1.12 & 1.27 & -1.51 & -1.51 & -1.41 & -1.64 \\ RX
For &    38 &  0.597 & 1.22 & 1.61 & 1.77 & -1.26 & -1.10 & &
\\ SS For &    50 &  0.495 & 1.14 & 1.50 & 1.65 & -1.35 & -1.37 &
&       \\ RR Gem &    39 &  0.397 & 1.18 & 1.60 & 1.68 & -0.35 &
-0.25 & -0.55 & -0.58 \\ SZ Gem &    13 &  0.501 & 1.20 & 1.50 &
1.62 & -1.81 & -1.64 &       &       \\ RW Gru & 45 &  0.550 &
1.06 & 1.41 & 1.56 & -2.00 & -1.93 &       &
\\ SV Hya &    50 &  0.479 & 1.23 & 1.58 & 1.73 & -1.70 & -1.67 &
&       \\ SZ Hya &    23 &  0.537 & 1.24 & 1.60 & 1.67 & -0.78 &
-0.49 &       &       \\ WZ Hya &    16 &  0.538 & 0.95 & 1.25 &
1.37 & -1.30 & -1.31 &       &       \\ DG Hya &    16 &  0.430 &
0.98 & 1.28 & 1.40 & -1.42 & -1.38 &       &       \\ FY Hya & 50
&  0.637 & 1.04 & 1.34 & 1.57 & -2.33 & -2.25 &       &
\\ V Ind &     24 &  0.480 & 1.22 & 1.57 & 1.74 & -1.50 & -1.46 &
&       \\ SS Leo &    48 &  0.626 & 1.12 & 1.44 & 1.59 & -1.83 &
-1.68 & -1.73 & -1.73 \\ VY Lib &    49 &  0.534 & 0.96 & 1.29 &
1.43 & -1.32 & -1.6 &       &       \\ RR Lyr &    10 &  0.567 &
0.48 & 0.63 & 0.68 & -1.37 & -1.45 & -1.37 & -1.38 \\ KX Lyr & 31
&  0.441 & 1.02 & 1.38 & 1.50 & -0.46 & -0.36 &       &
\\ RV Oct &    50 &  0.571 & 1.10 & 1.44 & 1.56 & -1.34 & -1.39 &
&       \\ UV Oct &    50 &  0.543 & 1.01 & 1.33 & 1.46 & -1.61 &
-1.66 &       &       \\ ST Oph &    50 &  0.450 & 1.25 & 1.62 &
1.75 & -1.30 & -1.22 &       & -1.41 \\ V445 Oph &  50 &  0.397 &
0.86 & 1.20 & 1.29 & -0.23 & -0.26 & -0.30 & -0.42 \\ V452 Oph &
48 &  0.557 & 0.95 & 1.24 & 1.42 & -1.72 & -1.73 &       &
\\ V455 Oph &  49 &  0.454 & 0.83 & 1.11 & 1.26 & -1.42 & -1.63 &
&       \\ V816 Oph &  13 &  0.381 & 1.31 & 1.67 & 1.83 & -0.06 &
-0.31 &       &       \\ TY Pav &    50 &  0.710 & 0.86 & 1.11 &
1.27 & -2.31 & -2.1 &       &       \\ DN Pav &    50 &  0.468 &
1.24 & 1.63 & 1.71 & -1.54 & -1.69 &       &       \\ VV Peg & 22
&  0.488 & 1.11 & 1.44 & 1.62 & -1.88 & -1.73 & -1.71 & -1.73
\\ AV Peg &    22 &  0.390 & 0.96 & 1.34 & 1.44 & -0.14 & -0.08 &
-0.36 & -0.23 \\ BH Peg &    50 &  0.641 & 0.64 & 0.87 & 0.95 &
-1.38 & -1.16 & -1.17 & -1.25 \\ U Pic &     25 &  0.440 & 1.16 &
1.53 & 1.62 & -0.73 & -0.64 &       &       \\ BB Pup &    42 &
0.480 & 0.88 & 1.21 & 1.30 & -0.57 & -0.65 &       &       \\ HH
Pup &    47 &  0.391 & 1.25 & 1.65 & 1.78 & -0.69 & -0.58 & &
\\ V440 Sgr&   30 &  0.477 & 1.16 & 1.50 & 1.63 & -1.47 & -1.21 &
&       \\ U Scl &     34 &  0.493 & 1.22 & 1.55 & 1.70 & -1.25 &
-1.21 &       &       \\ RU Scl &    34 &  0.493 & 1.22 & 1.55 &
1.70 & -1.25 & -1.21 &       &       \\ VY Ser & 50 &  0.714 &
0.69 & 0.94 & 1.05 & -1.82 & -1.53 & -1.83 & -1.08
\\ AN Ser &    50 &  0.522 & 0.97 & 1.30 & 1.39 & -0.04 & -0.11 &
-0.39 & -0.29 \\ AR Ser &    34 &  0.575 & 0.87 & 1.13 & 1.29 &
-1.78 & -1.64 & -1.48 &       \\ AV Ser &    50 &  0.488 & 1.10 &
1.44 & 1.59 & -1.20 & -1.37 &       &       \\ RW Tra &    50 &
0.374 & 0.75 & 1.08 & 1.21 & -0.07 & -0.14 &       &       \\ W
Tuc &     41 &  0.642 & 0.95 & 1.27 & 1.42 & -1.64 & -1.85 & &
\\ YY Tuc &    49 &  0.635 & 1.16 & 1.50 & 1.67 & -1.82 & -1.85 &
&       \\ RV UMa &    19 &  0.468 & 1.09 & 1.42 & 1.50 & -1.19 &
-1 & -0.96 & -1.07 \\ TU UMa &    12 &  0.558 & 0.89 & 1.17 & 1.27
& -1.44 & -1.51 & -1.37 & -1.59 \\ FS Vel & 19 &  0.476 & 0.93 &
1.23 & 1.34 & -1.17 & -1.19 &       &
\\ UU Vir &    46 &  0.476 & 1.11 & 1.46 & 1.58 & -0.82 & -0.93 &
-0.95 & -0.70 \\ AM Vir &    24 &  0.615 & 0.64 & 0.86 & 1.00 &
-1.45 & -1.18 &       &       \\ AV Vir &    49 &  0.657 & 0.75 &
1.00 & 1.13 & -1.32 & -1.46 & -1.26 & -1.31 \\
\hline
\noalign{\smallskip}
 \end{tabular}
\end{table}

\begin{table}[h]
\caption[ ]{}
\begin{tabular}{lccccccccc}
\hline \noalign{\smallskip}
 Var & nph & Per & $A_{V}$&$A_{B}$& $A_{U}$&$[Fe/H]_{L}$
&$[Fe/H]_{UBV}$&$[Fe/H]_{SKK}$&$[Fe/H]_{LS}$\\
 \hline
\noalign{\smallskip}

 AT And  &   14  & 0.617  & 0.48  &
0.64  & 0.69  & -0.81  & -1.19  & -1.06  \\ S Ara   &   50  &
0.452  & 1.26  & 1.62  & 1.75  & -1.41  &        & -0.83  \\ RU
Boo  &   19  & 0.493  & 1.10  & 1.51  & 1.85  & -1.49  &        &
\\ AA CMi  &   11  & 0.476  & 0.98  & 1.34  & 1.44  & -0.30  &
&        \\ UU Cet  &   47  & 0.606  & 0.65  & 0.86  & 0.99  &
-1.35  & -1.88  & -1.04  \\ Z Com   &   17  & 0.547  & 1.11  &
1.48  & 1.69  & -1.52  &        &        \\ UY Cyg  &   27  &
0.561  & 0.83  & 1.11  & 1.22  & -0.93  & -0.95  & -0.86  \\ VX
Her  &   50  & 0.455  & 1.29  & 1.63  & 1.76  & -1.44  & -1.37  &
-1.52  \\ AR Her  &   22  & 0.470  & 0.92  & 1.15  & 1.27  & -1.83
&        & -1.41  \\ CZ Lac  &   18  & 0.432  & 0.46  & 8.87  &
1.11  &        & -0.54  & -0.52  \\ TT Lyn  &   23  & 0.597  &
0.69  & 0.93  & 1.01  & -0.98  & -1.84  & -1.54  \\ FN Lyr  &   27
& 0.527  & 1.25  & 1.61  & 1.94  & -1.52  &        &        \\ RV
Phe  &   48  & 0.596  & 0.79  & 1.05  & 1.25  & -1.36  &        &
\\ XX Pup  &   20  & 0.517  & 1.23  & 1.56  & 1.71  & -1.40  &
&        \\ V675 Zgr&   48  & 0.642  & 0.92  & 1.21  & 1.37  &
-1.92  &        &        \\ V494 Sco&   50  & 0.427  & 0.97  &
1.27  & 1.43  & -1.67  &        &        \\ V690 Sco&   50  &
0.492  & 1.08  & 1.43  & 1.55  & -0.86  &        &        \\ AM
TUc  &   50  & 0.406  & 0.44  & 0.58  & 0.69  & -2.54  &        &
\\ AF Vel  &   50  & 0.527  & 0.91  & 1.23  & 1.36  & -1.54  &
&        \\ BB Vir  &   15  & 0.471  & 0.68  & 0.82  & 0.89  & & &
-1.57  \\ BN Vul  &   15  & 0.594  & 0.57  & 0.71  & 0.78  & &
-1.61  & -1.53  \\
 \hline \noalign{\smallskip}
 \end{tabular}
\end{table}

\begin{table}[h]
\caption[ ]{}
\begin{tabular}{lccccccccc}
\hline \noalign{\smallskip}
 Var & nph & Per & $A_{V}$&$A_{B}$&$A_{U}$&$[Fe/H]_{L}$
&$[Fe/H]_{UBV}$&$[Fe/H]_{SKK}$&$[Fe/H]_{LS}$\\ \hline
\noalign{\smallskip}

TY Aps& 50 &    0.502&  1.01&   1.30&   1.48&   -1.21&  -1.73& &
\\ BR Aqr& 50 &    0.482&  1.13&   1.48&   1.55&   -0.84& -1.30&
&  -0.88\\ UY Boo& 50 &    0.651&  1.10&   1.42& 1.60&   -2.49&
-2.12&  -2.40&  -2.53\\ RZ Cet& 50 &    0.510& 0.93&   1.22& 1.38&
-1.50&  -1.90&  -1.47&  -1.34\\ ST Leo& 48 &    0.478& 1.24& 1.62&
1.75&   -1.29&  -1.69&       & -1.29\\ TV Leo& 49 & 0.673&  1.14&
1.47&   1.63&   -1.97& -1.47&  -2.07& -2.12\\ RY Psc& 50 & 0.530&
0.79&   1.04& 1.20&   -1.39& -2.26&  -1.52& -1.77\\ AT Ser& 50 &
0.745& 0.87&   1.14& 1.26&   -2.05& -1.55&  -1.93& -2.00\\
 \hline
\noalign{\smallskip}
 \end{tabular}
\end{table}

{}

\end{document}